\documentclass[]{aastex631}
\shorttitle{Double Mode Cepheids from the ZTF Survey}
\shortauthors{V. Shah, X. Chen, and R. de Grijs}
\graphicspath{ {Figures/} }
\begin{document}

\title{Double Mode Cepheids from the Zwicky Transient Facility Survey}

\correspondingauthor{Richard de Grijs, Xiaodian Chen}
\email{richard.de-grijs@mq.edu.au, chenxiaodian@nao.cas.cn}

\author{Vishwangi Shah}
\affiliation{Birla Institute of Technology and Science, Pilani, Hyderabad Campus, India}
\affiliation{School of Mathematical and Physical Sciences, Macquarie University, Balaclava Road, Sydney, NSW 2109, Australia}
\affiliation{Department of Physics, McGill University, 3600 rue University, Montr\'{e}al, QC H3A 2T8, Canada}
\affiliation{McGill Space Institute, McGill University, 3550 rue University, Montr\'{e}al, QC H3A 2A7, Canada}

\author{Xiaodian Chen}
\affiliation{CAS Key Laboratory of Optical Astronomy, National Astronomical Observatories, Chinese Academy of Sciences, Beijing 100101, China}

\author{Richard de Grijs}
\affiliation{School of Mathematical and Physical Sciences, Macquarie University, Balaclava Road, Sydney, NSW 2109, Australia}
\affiliation{Research Centre for Astronomy, Astrophysics and Astrophotonics, Macquarie University, Balaclava Road, Sydney, NSW 2109, Australia}

%% Note that the \and command from previous versions of AASTeX is now
%% depreciated in this version as it is no longer necessary. AASTeX 
%% automatically takes care of all commas and "and"s between authors names.

%% AASTeX 6.31 has the new \collaboration and \nocollaboration commands to
%% provide the collaboration status of a group of authors. These commands 
%% can be used either before or after the list of corresponding authors. The
%% argument for \collaboration is the collaboration identifier. Authors are
%% encouraged to surround collaboration identifiers with ()s. The 
%% \nocollaboration command takes no argument and exists to indicate that
%% the nearby authors are not part of surrounding collaborations.

%% Mark off the abstract in the ``abstract'' environment. 
\begin{abstract}
Multi-mode Cepheids pulsate simultaneously in more than one mode of oscillation. They provide an independent means to test stellar models and pulsation theories. They can also be used to derive metallicities. In recent years, the number of known multi-mode Cepheids has increased dramatically with the discovery of a large number of Galactic double-mode Cepheids. To date, 209 double-mode Cepheids have been detected in the Galactic bulge and disk, mostly based on the Optical Gravitational Lensing Experiment's (OGLE) catalog. In this paper, we conduct a comprehensive search for double-mode Cepheids in the northern sky based on Zwicky Transient Facility Data Release 5. We found 72 such objects in the Milky Way. The periods of the 30 sample objects already included in the OGLE catalog show excellent agreement with the OGLE periods. The period ratios of our new Cepheids are consistent with those of known double-mode Cepheids, as evidenced by their loci in the so-called `Petersen diagram'. Compared with OGLE, the completeness of our double-mode Cepheid sample is around 71\%. The much improved temporal sampling of the Zwicky Transient Facility offers significant scope to find more double-mode Cepheids, especially at the distribution's short-period end.
\end{abstract}

%% Keywords should appear after the \end{abstract} command. 
%% The AAS Journals now uses Unified Astronomy Thesaurus concepts:
%% https://astrothesaurus.org
%% You will be asked to selected these concepts during the submission process
%% but this old "keyword" functionality is maintained in case authors want
%% to include these concepts in their preprints.

%% From the front matter, we move on to the body of the paper.
%% Sections are demarcated by \section and \subsection, respectively.
%% Observe the use of the LaTeX \label
%% command after the \subsection to give a symbolic KEY to the
%% subsection for cross-referencing in a \ref command.
%% You can use LaTeX's \ref and \label commands to keep track of
%% cross-references to sections, equations, tables, and figures.
%% That way, if you change the order of any elements, LaTeX will
%% automatically renumber them.
%%
%% We recommend that authors also use the natbib \citep
%% and \citet commands to identify citations.  The citations are
%% tied to the reference list via symbolic KEYs. The KEY corresponds
%% to the KEY in the \bibitem in the reference list below. 

\section{Introduction} \label{sec:intro}

Classical Cepheids are pulsating yellow giant or supergiant stars that exhibit periodic variations in their surface temperatures and luminosities. They are extremely bright and characterized by luminosities equivalent to 500--$3 \times 10^4 L_\odot$  which makes them observable and distinguishable out to distances of up to $\sim$30 Mpc or more \citep{book1}. Moreover, they exhibit a tight period--luminosity relationship, which makes them crucial contributors to the cosmic distance ladder. 

%Cepheid variables can be broadly classified into Types I and II. Type I Cepheids are young and have high(er) metallicities, as well as characteristic absorption and emission lines that can be used to identify them. They are luminous and can thus be used to probe great distances. By contrast, Type II Cepheids are old, have low(er) metallicities, and are less luminous than Type I Cepheids. Type II Cepheids are often used to derive distances to galactic centers, galactic disks, and globular clusters \citep{book1}. Type I Cepheids obey power-law period--luminosity relations (PLRs), which in turn allow precise distance measurements. PLRs are different for different galaxies, since they are affected by variations and differences in metallicities, star-formation histories, and extinction properties. This hence calls for a more in-depth understanding of the internal structure of Cepheids \citep{paper5}.

Cepheids may exhibit a range of oscillation modes, including in their fundamental (F) mode, as well as in their first and/or second overtones (1O/2O). They sometimes oscillate in more than one mode at the same time, thus producing double- and even triple-mode Cepheids \citep{paper3}. The period ratio for Cepheids pulsating in both the first- and second-overtone modes is $P_{21} = P_2/P_1 \simeq 0.80$, while for Cepheids pulsating in both the fundamental and first overtone modes, $P_{10} = P_1 /P_0 \simeq 0.72$ \citep{2020AcA....70..101S}. The period ratios of the F/1O double-mode Cepheids in the Milky Way, the Large Magellanic Cloud (LMC), and the Small Magellanic Cloud (SMC) are different, that is, Cepheids in those galaxies occupy different regions in the so-called `Petersen diagram' \citep{petersen1973}, as a result of their different metallicities \citep{2007ApJ...660..723B}. The Petersen diagram covers the parameter space defined by the ratio of the shorter to the longer period versus the logarithm of the longer period \citep{soszynski2017, soszynski2020, udalski2018}. 

Multi-mode Cepheids play an important role in studies of stellar evolution, since they probe the structure of the stellar envelope and can be used to test pulsation theories \citep{paper6, 2010A&A...524A..40S}. The period ratios of multi-mode Cepheids can be used to determine their masses and radii without the need to know their luminosities, effective temperatures, or surface gravities \citep{paper7,1992ApJ...385..685M}. On the other hand, $P_{10}$ of multi-mode Cepheids can be used to determine their metallicities. \citet{2007A&A...473..579S} and \citet{2016MNRAS.460.2077K} calibrated the relationship between $P_1/P_0$ and [Fe/H] using high-resolution spectroscopy. Those authors used it to study the metallicity distributions of the young stellar populations in the Magellanic Clouds. \citet{2018A&A...618A.160L} applied this same calibration to the F/1O double-mode Cepheids discovered by {\sl Gaia} \citep{2019A&A...622A..60C} to derive the metallicity gradient in the Galactic disk.

Extragalactic multi-mode Cepheids have been studied by many authors, e.g., in M31 \citep{2013ApJ...777...35L,2013ApJ...778..147P}, M33 \citep{2006ApJ...653L.101B}, the LMC \citep{1999ApJ...511..185A, 2008AcA....58..153S, 2009A&A...495..249M}, and the SMC \citep{2009A&A...495..249M, 2010AcA....60...17S}. Multi-mode Cepheids in the Milky Way are more valuable, but their number included only a few dozen objects prior to 2018. Since then, the Optical Gravitational Lensing Experiment's (OGLE) catalog \citep{soszynski2017, soszynski2020, udalski2018} now contains 1973 Cepheids located in the Milky Way's bulge and disk. Approximately 11\% (209) of these Cepheids are F/1O and 1O/2O double-mode Cepheids \citep{2020AcA....70..101S}. \citet{2018AcA....68..341J} found 15 multi-mode Cepheids based on the ASAS-SN variables catalog \citep{2018MNRAS.477.3145J, 2020MNRAS.491...13J}. NASA's Transiting Exoplanet Survey Satellite (TESS) data is suitable to find multi-mode Cepheids with smaller amplitudes \citep{2021ApJS..253...11P}. \citet{paper1} found about 700 new Cepheids based on Zwicky Transient Facility (ZTF) Data Release 2 (DR2). {\sl Gaia} has also been used to detect new Cepheids \citep{2019A&A...625A..14R}, as have the Asteroid Terrestrial-impact Last Alert System \citep{2018AJ....156..241H} and the Wide-field Infrared Survey Explorer \citep{2018ApJS..237...28C}. Combining these Cepheid catalogs, here we report on a comprehensive search for double-mode Cepheids based on ZTF DR5.\footnote{https://www.ztf.caltech.edu/page/dr5} Section \ref{sec:datanandmethods} describes the data and methods used in our analysis. Section \ref{sec:results} outlines the results obtained, which we discuss in Section \ref{sec:disc}. We conclude the paper in Section \ref{sec:concl}.

\section{Data and Methods} \label{sec:datanandmethods}

ZTF is a robotic optical time-domain survey that uses the 48-inch Samuel Oschin Telescope at Palomar Observatory \citep{2019PASP..131a8003M}. It has a 47 deg$^2$ field of view, which enables observation of the entire visible northern sky. It provides photometry in the $g$ and $r$ bands. ZTF's main science goal is the detailed study of variable and transient astrophysical sources \citep{graham2019}. For public survey purposes, the entire sky visible from Palomar Observatory is observed every three nights, whereas the visible Galactic plane is covered every night. Over a period of three years, some $10^9$ sources have been observed 300--500 times each. ZTF DR5 contains data acquired between 2018 March and 2021 January. 

Combining the Cepheid catalogs referred to in Section \ref{sec:intro}, we found 1436 classical Cepheids in the Milky Way that were well sampled in ZTF DR5. There were instances where a particular source was observed more than once and thus had different observation IDs (oids; see below). We identified the different observations corresponding to every single source and merged them. Bad observational epochs were removed from the light curves by adopting {\tt catflags} $< 32768$. Some light curves contained clusters of data points with a cadence of less than 0.001 days. We also removed those data points, since they tended to skew our period analysis toward incorrect periods by weighing those short-cadence points disproportionately. Finally, we excluded Cepheids for which we had access to fewer than 20 observational epochs. 

We performed period analysis of our sample sources based on Lomb--Scargle periodograms \citep{lomb1976, scargle1982}. The Lomb--Scargle periodogram is the most suitable technique to obtain periods based on unevenly spaced data. We used the python function {\tt astropy.timeseries.LombScargle} to derive Lomb--Scargle periodograms from the Cepheids' light curves. This function returns a power spectrum in the form of an array of frequencies and their corresponding powers. We set the minimum and maximum frequencies  of the power spectrum to 0.00001 day$^{-1}$ and 10 day$^{-1}$, respectively, since ZTF periods are found within this frequency range \citep[e.g.,][]{paper1}. We also set the samples-per-peak to 40 so as to make sure that our grid sampled each peak sufficiently well. The frequency corresponding to the highest peak in the power spectrum was recorded and its reciprocal value was adopted as the primary period ($P_1$). The light curves were pre-whitened with $P_1$, i.e., the Fourier peak corresponding to $P_1$ was removed, and a second power spectrum was obtained to find any secondary periods ($P_2$). For most sample objects, $P_1 > P_2$, although a small number of objects resulted in $P_2 > P_1$. We recorded such cases as well. 

The uncertainties in our periods were expressed using their `false alarm probabilities' (FAPs). The FAP is a means to quantify the significance of a periodogram peak. It quantifies the probability that a data set with no periodic signal may yield a peak of similar magnitude because of a coincidental alignment of random errors \citep{vanderplas2018}. We excluded all sources that had FAPs for the primary and/or secondary periods $>$ 0.001. We visually analyzed the phase-folded light curves of all our sources and excluded those sources that did not show clear periodic trends, those that had folded light-curve shapes similar to those of eclipsing binaries, those where $P_2$ was an alias (multiple) of $P_1$, those that were demonstrably aliased because of the daily observational cadence, and those whose aliased frequencies ($f$) coincided with the equivalent combination of $P_1$ and $P_2$, i.e., where  $f \equiv f_{P_1} \pm f_{\rm aliased} \simeq f_{P_2}$.

The FAP for the highest peak (FAP1) was recorded. We folded the light curve with $P_1$ and fitted it with a tenth-order Fourier function. The best-fitting function was subtracted from the light curve, returning residual magnitudes at each observational epoch. This same procedure was repeated for the residual data to obtain $P_2$ and the corresponding FAP2.

Valid $P_2$ values satisfied $0.65 < P_2/P_1 < 0.85$ if $P_2 < P_1$ (or $0.65 < P_1/P_2 < 0.85$ if $P_1 < P_2$). These limits were adopted based on the distinct loci occupied by the Milky Way's F/1O and 1O/2O classical Cepheids in the Petersen diagram. We applied our periodogram analysis to both the $g$ and $r$ data, and we adopted a source as a candidate double-mode Cepheid if it satisfied our selection criteria in at least one passband. For almost all sources in our data set, the resulting periods, both $P_1$ and $P_2$, were identical in both filters. In a small number of cases, we obtained multiple possible values for the secondary periods. This condition applied to three objects, covered by six different oids. Specifically, 
\begin{enumerate}
    \item oids 539103300012020 and 539203300061346 referred to the same object (ZTFJ185513.28+081813.6, $P_1=1.86719$ d) but returned 1.38460 d and 1.34256 d as the $P_2$ values (with $P_1 > P_2$) in the $g$ and $r$ bands, respectively. We chose the period 1.34256 d as the real $P_2$ value, because it had a significantly smaller FAP2.
    \item oids 461102300023057 and 461202300006319 (ZTFJ065417.34+012748.2, $P_1=0.64064$ d) yielded three different $P_2$ values, 0.52908 d and 0.47214 d (for $P_1 > P_2$) in the $g$ and $r$ bands, respectively, and additionally 0.89663 d (for $P_2 > P_1$) in the $r$ band. This Cepheid is unlikely a double-mode Cepheid, because the amplitude of the second period is lower than expected for double-mode Cepheids ($\sim$0.03 mag) and the period ratio is inconsistent with other double-mode Cepheids. 
    \item oids 461110100019729 and 461210100026719 (ZTFJ065759.85+053444.9, $P_1=0.97814$ d) returned three different light-curve fits. For $P_1 > P_2$, the corresponding $P_2$ value was 0.79866 d in the $g$ and $r$ bands, but for $P_2 > P_1$ the $g$-band value of $P_2$ was 1.33215 d. The $P_1$ value for all three cases was 0.97814 d. We found that this is a triple-mode Cepheid with a fundamental period of 1.33215 d, a first-overtone period of 0.97814 d, and a second-overtone period of 0.79866 d, since the two period ratios resulting from these choices agree well with those of other double-mode Cepheids. 
    \end{enumerate}

\begin{figure}[h]
	    \centering
	    %\caption{Folded ZTF lightcurves with their Fourier fits}
	    \includegraphics[scale=0.5]{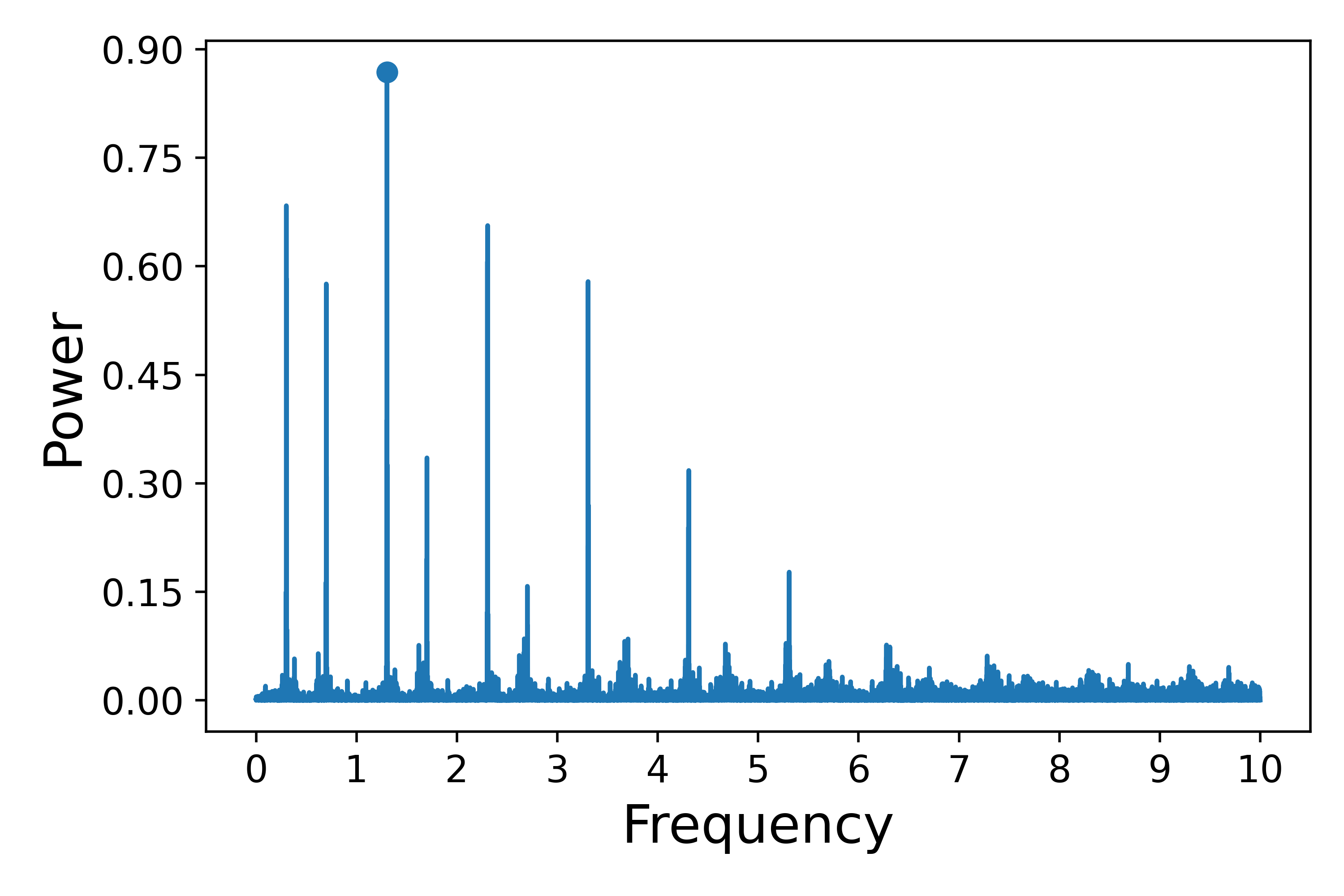}
	    \includegraphics[scale=0.5]{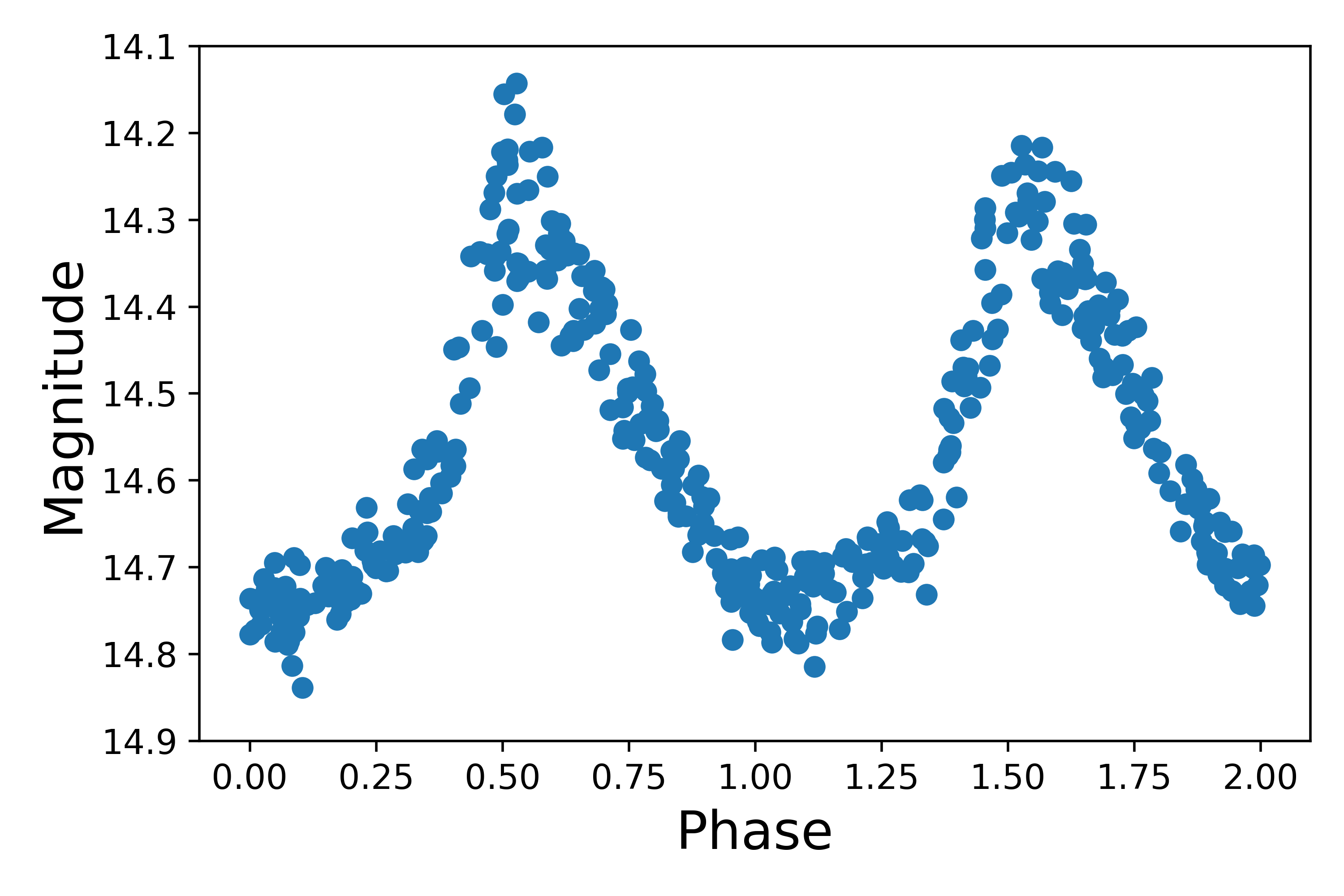}
	   \includegraphics[scale=0.5]{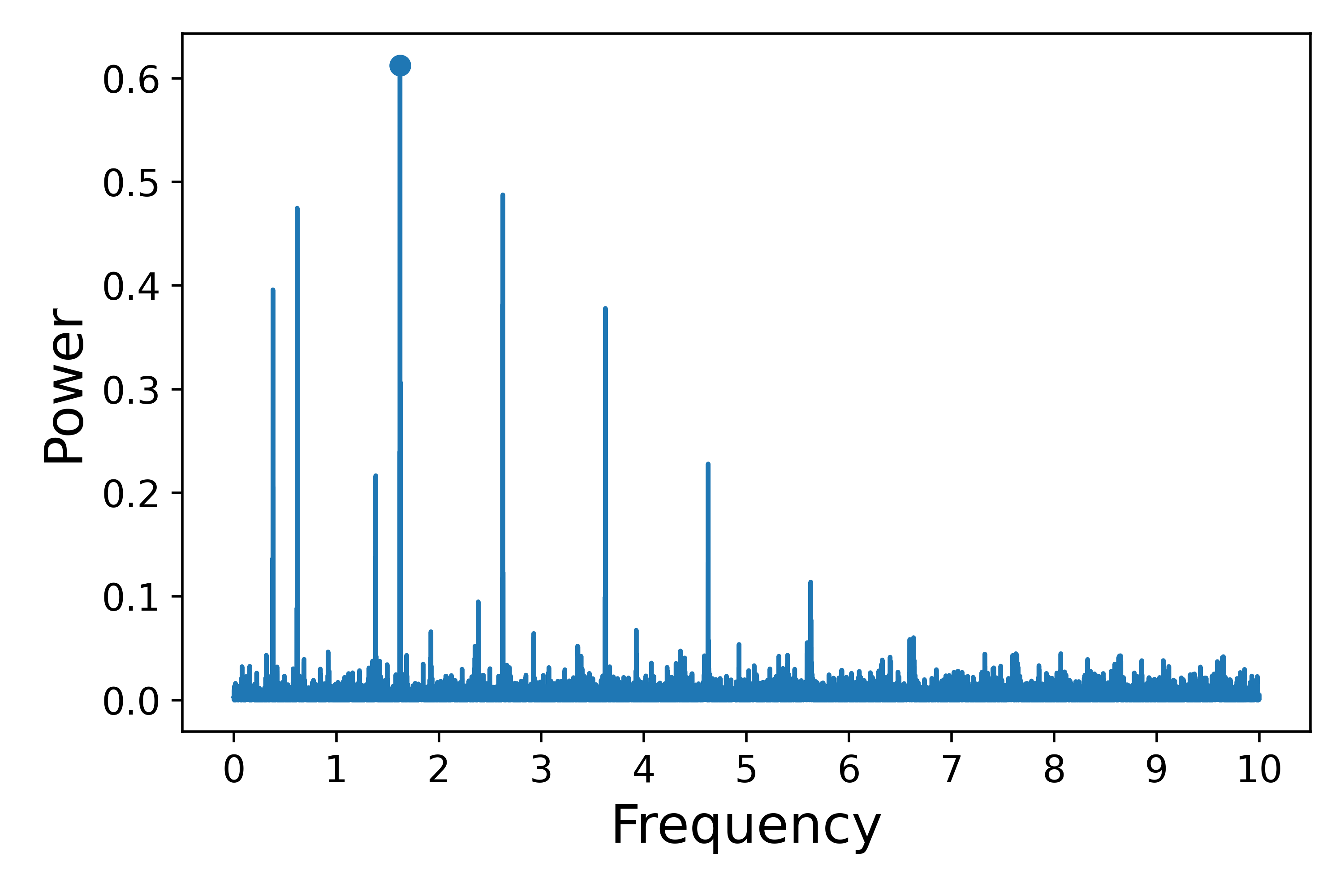}
	   \includegraphics[scale=0.5]{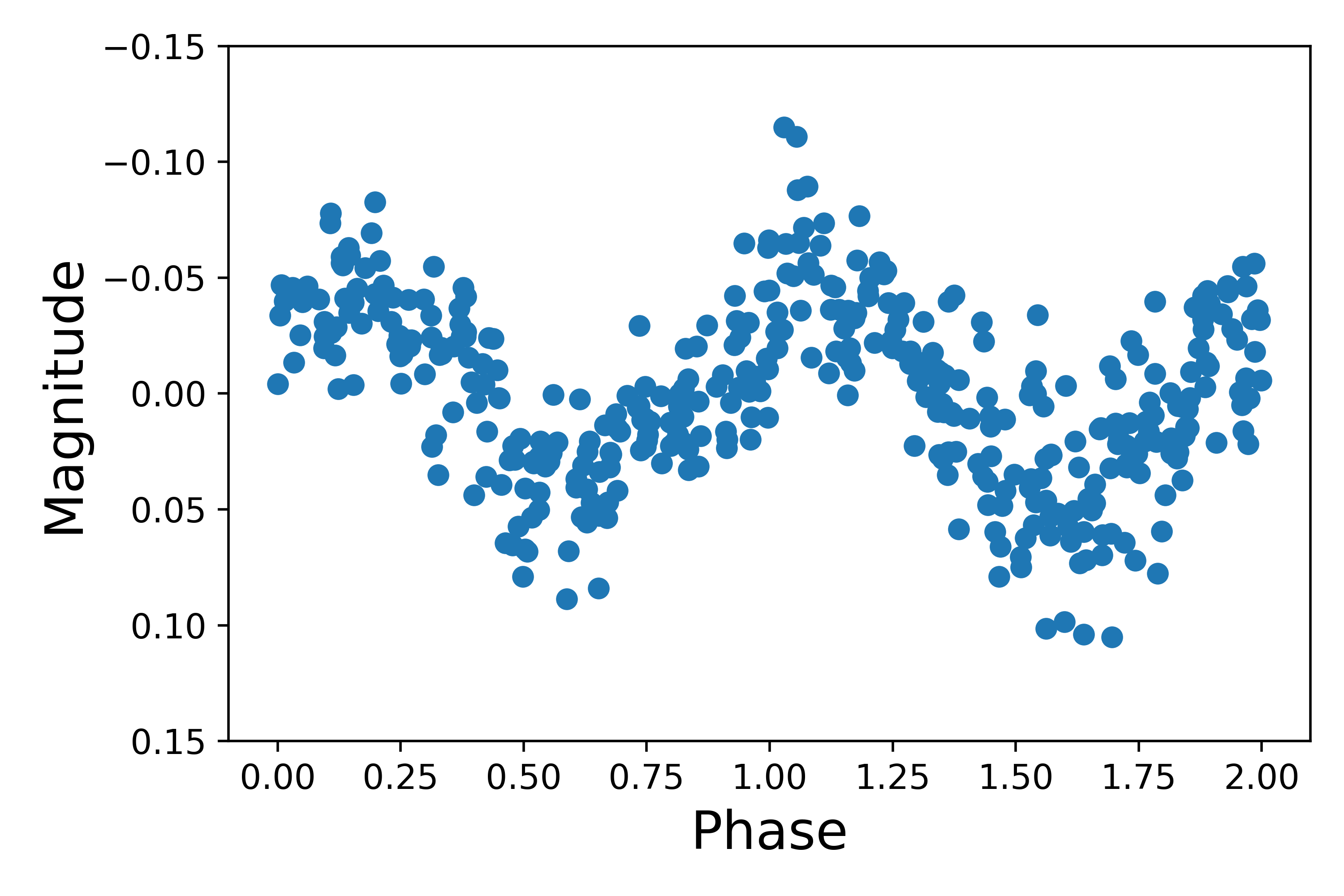}
	    \caption{Power spectra and folded light curves for the double-mode Cepheid ZTFJ195955.52+363159.9. (Top left) Power spectrum of the original data with the highest peak ($P_1$) marked; (top right) Original light curve phased with $P_1$; (bottom left) Power spectrum after pre-whitening of the original light curve with $P_1$, with the highest peak ($P_2$) marked; (bottom right) Pre-whitened light curve phased with $P_2$ after removing the power spectrum peaks corresponding to $2 f_{P_2}$, $3 f_{P_2}$, $(f_{P_1} + f_{P_2})$, and $(f_{P_2} - f_{P_1}$).}
	    \label{fig:1}
	\end{figure}

\begin{figure}[h]
	    \centering
	    %\caption{Folded ZTF lightcurves with their Fourier fits}
	    \includegraphics[scale=0.5]{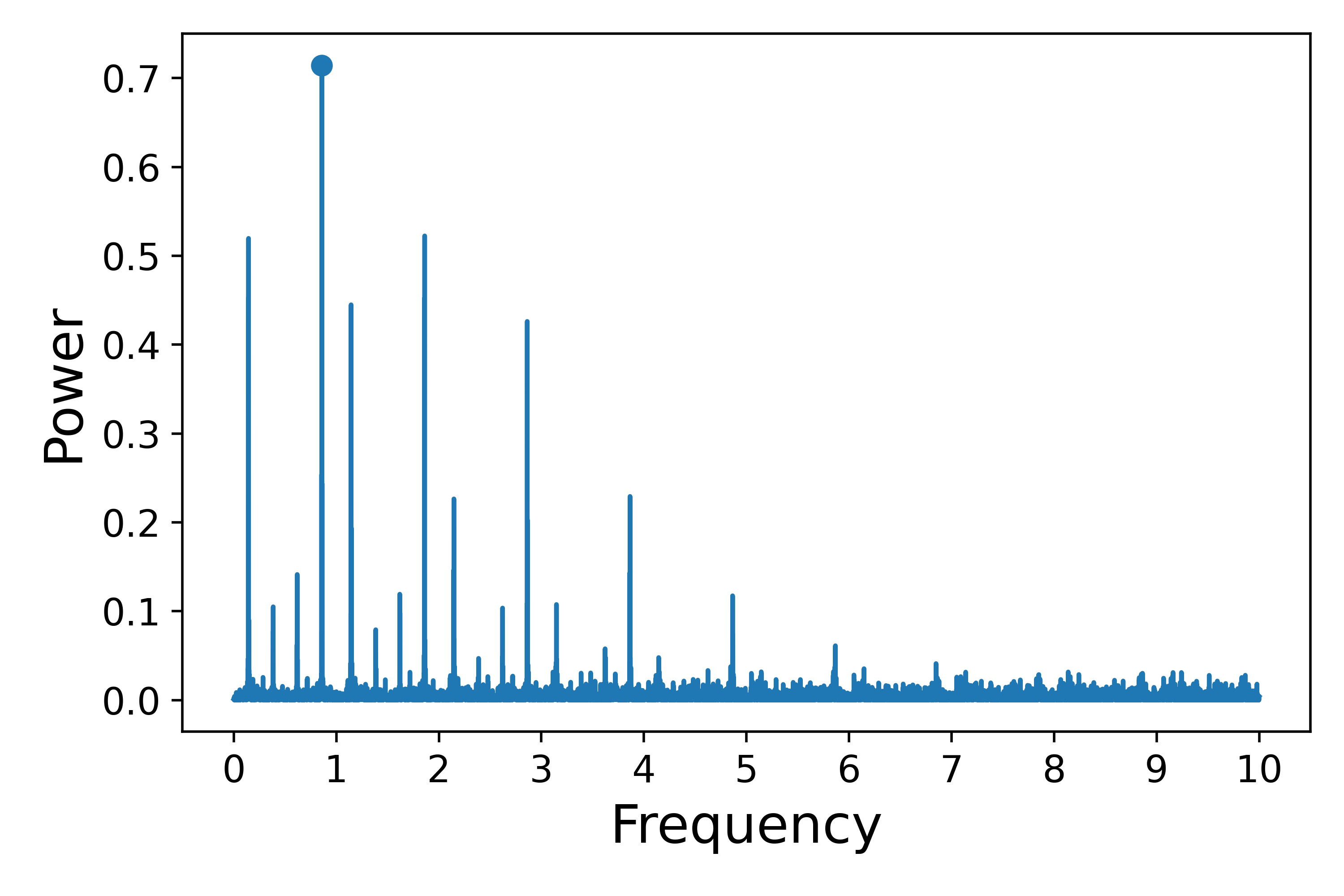}
	    \includegraphics[scale=0.5]{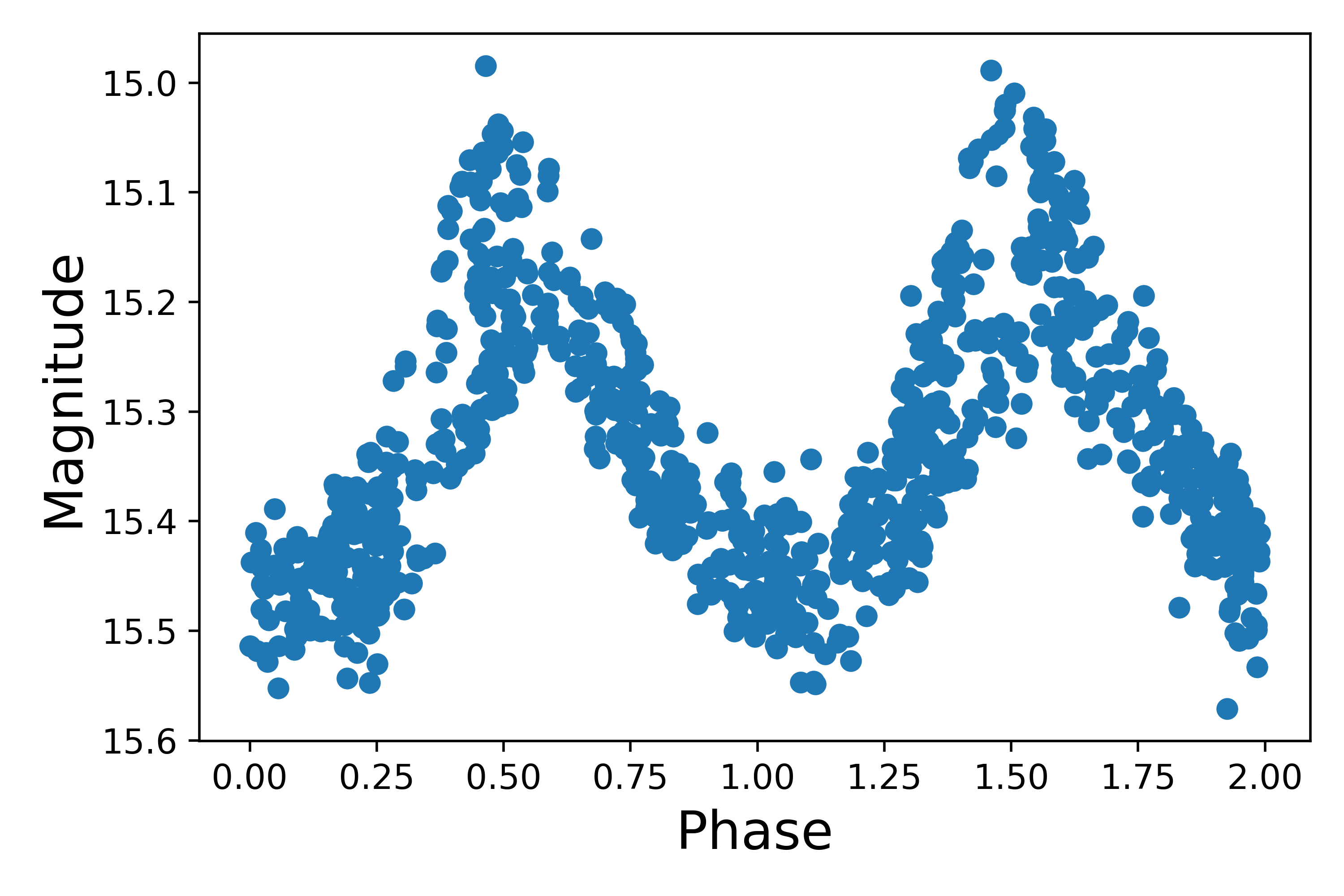}
	   \includegraphics[scale=0.5]{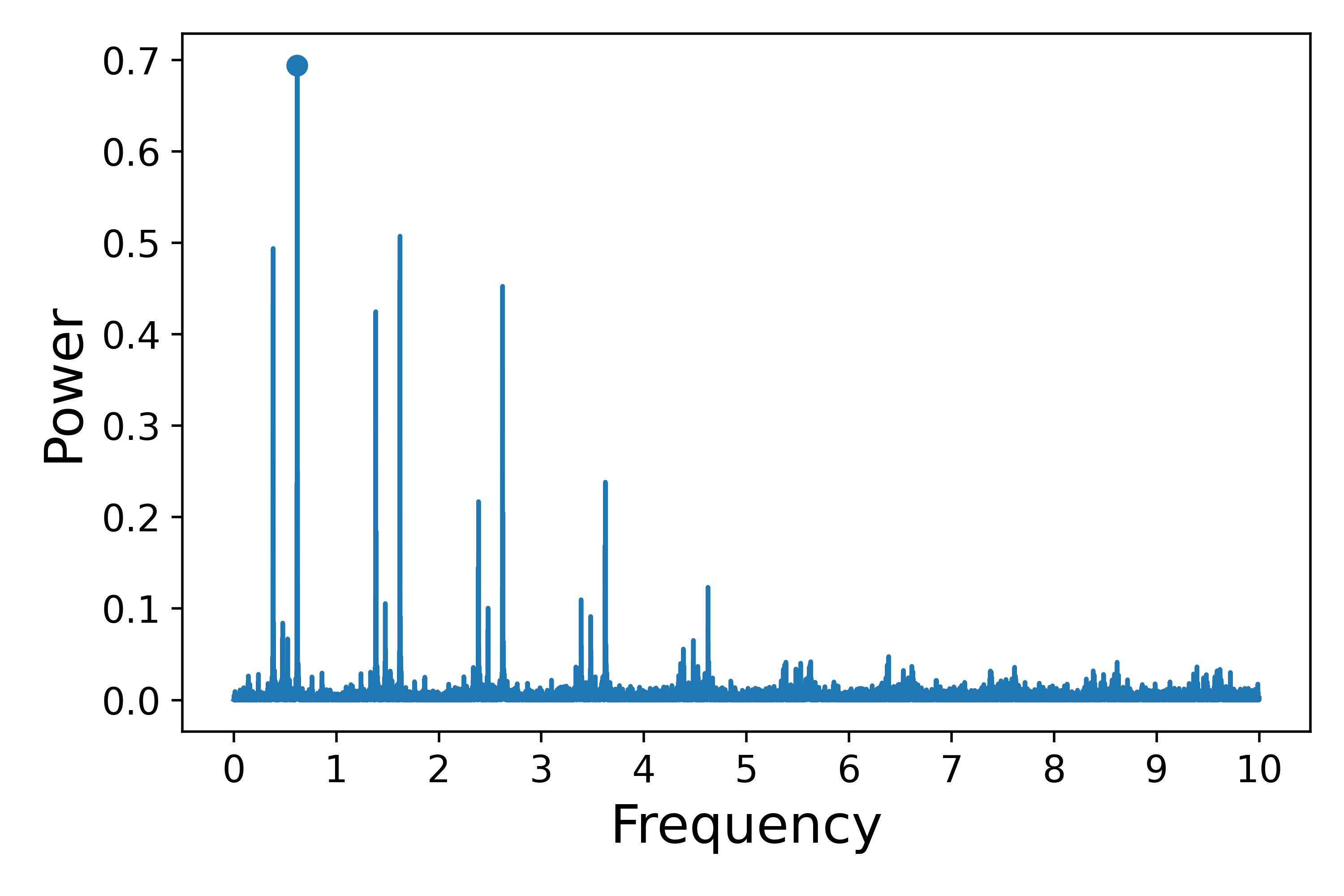}
	   \includegraphics[scale=0.5]{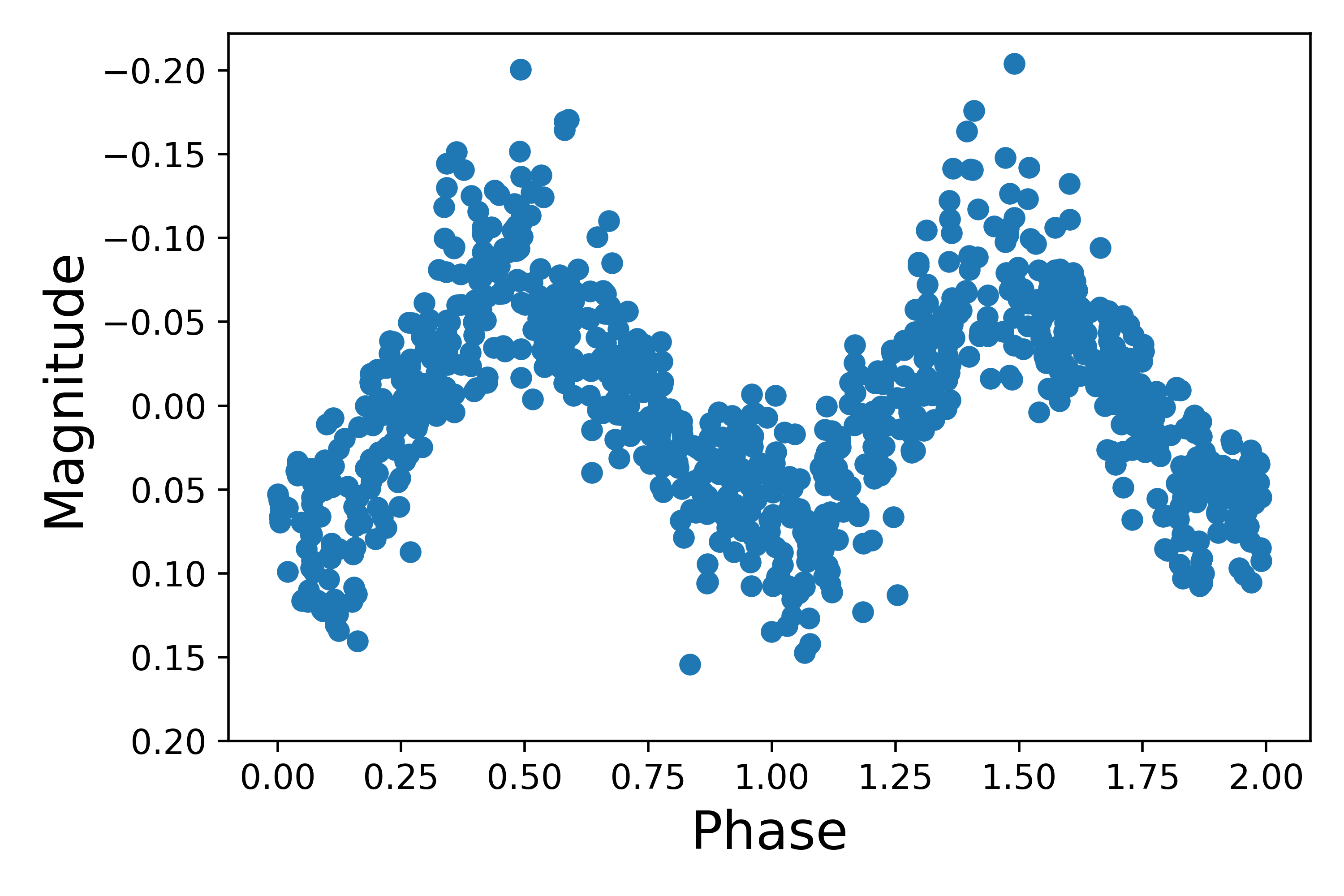}
	    \caption{As Figure \ref{fig:1}, but for the double-mode Cepheid ZTFJ210811.91+460046.7.
	    \label{fig:2}}
	\end{figure}

Our final tally of double-mode Cepheid candidates (see Table \ref{tab:new_DMC}) includes 72 objects. Of those, 49 sources were detected in both the $g$ and $r$ bands, where they yielded the same $P_1$ and $P_2$ values, 14 sources were detected only in the $g$ band, and nine sources were detected only in the $r$ band. Figures \ref{fig:1} and \ref{fig:2} show examples of double-mode Cepheids identified in our analysis.

We assigned modes (F/1O or 1O/2O) to the double-mode Cepheids based on their period ratios. The OGLE F/1O and 1O/2O Cepheids occupy specific loci in the Petersen diagram, such that their period ratios lie between 0.65 and 0.85 \citep{2020AcA....70..101S}. For the Galactic double-mode Cepheids included in the OGLE catalog \citep{soszynski2017, soszynski2020, udalski2018}, sources with period ratios $>$ 0.769 were classified as 1O/2O and those with period ratios $<$ 0.769 were classified as F/1O double-mode Cepheids. We used the same selection criteria to classify our ZTF-based sample of double-mode Cepheid candidates. 

\section{Results} \label{sec:results}

For our 72 double-mode Cepheid candidates, Table \ref{tab:new_DMC} lists their ZTF ID, right ascension (J2000), declination (J2000), $P_1$, FAP1, $P_2$, FAP2, period ratio, and mode. Thirty of these Cepheids were already included in the OGLE database. Table \ref{tab:OGLE_DMC} lists the OGLE periods as well as the periods resulting from our analysis of the sources in common. Our periods are in excellent agreement with the OGLE periods. Following a comparison with other double-mode Cepheid catalogs, we found that five double-mode Cepheids were detected in analyses based on the ASAS-SN database \citep{2018AcA....68..341J}. Sixty-one of our candidates were identified as double-mode Cepheid candidates in a recent paper based on all available databases \citep{2021AcA....71..205P}, while 11 double-mode Cepheid candidates are newly detected.

\begin{deluxetable*}{ccccccccc}
\tablenum{1}
\tablecaption{Double-mode Cepheids in ZTF DR5\label{tab:new_DMC}}
\tablewidth{0pt}
\tablehead{
\colhead{ZTF ID} & \colhead{R.A. (J2000)} & \colhead{Dec. (J2000)} & \colhead{$P_1$} & \colhead{FAP1} & \colhead{$P_2$} & \colhead{FAP2} & \colhead{Period Ratio} & \colhead{Modes}\\
\colhead{} & \colhead{(deg)} & \colhead{(deg)} & \colhead{(days)} & \colhead{} & \colhead{(days)} & \colhead{} & \colhead{} & \colhead{}
}
%\decimalcolnumbers
\startdata
ZTFJ213546.72+564559.3$^b$	& 323.9446686 &	   56.7664807 & 0.75803 &	3.02E$-$45  & 0.60960  & 1.56E$-$04  &	0.80419 & 1O/2O \\
ZTFJ035930.54+505514.2$^c$	&  59.8772594 &	   50.9206351 &	3.37295 &	7.41E$-$90  & 2.44433 & 2.74E$-$04  &	0.72469 & F/1O \\
ZTFJ044322.92+465704.3$^a$	&  70.8455148 &	   46.9512165 & 0.97555 &	8.32E$-$97  & 0.78273 & 5.21E$-$36  &	0.80235 & 1O/2O\\
ZTFJ054002.90+160503.3$^a$	&  85.0120902 &	   16.0842733 & 0.63082	& 4.91E$-$78	& 0.50916	& 2.05E$-$27 &	0.80714 & 1O/2O \\
ZTFJ055844.57+072750.2$^b$	&  89.6857337 &	    7.4639545 &	0.68818	& 6.12E$-$47 &	0.55460 &	7.52E$-$12 &	0.80589 & 1O/2O \\
ZTFJ073008.66$-$194457.4$^b$  & 112.5361077 &	$-$19.7492858 & 0.69966	&	1.11E$-$47	&	0.56276	&	5.22E$-$10	&	0.80433 & 1O/2O \\
ZTFJ074310.75$-$113457.5$^b$  & 115.7948270 &	$-$11.5826560 &	0.71538	&	5.96E$-$32	&	0.57740	&	4.15E$-$14	&	0.80713 &	1O/2O \\
%539203300061346 & 283.8050783 &     8.3037114 &	1.867 &	8.93E$-$67  & 1.343 & 3.85E$-$26  &	0.719 & F/1O - 141 \\
ZTFJ192918.18+220940.4$^c$  & 292.3257764 &    22.1612350 &	1.29991	& 2.88E$-$198	& 1.04274	& 3.67E$-$20 &	0.80216 &	F/1O \\
ZTFJ195955.52+363159.9$^a$	& 299.9813390 &	   36.5333247 &	0.76698	&	8.81E$-$147	&	0.61685	&	2.32E$-$63	&	0.80426 & 1O/2O \\
ZTFJ200505.47+311045.0$^a$	& 301.2728312 &	   31.1791667 &	1.88221	&	2.49E$-$98	&	1.50721	&	8.21E$-$18	&	0.80076 &	1O/2O \\
ZTFJ202946.50+374539.3$^a$  & 307.4437803 &	   37.7609270 &	4.29728	&	3.48E$-$105	&	2.99001	&	1.24E$-$45	&	0.69579 &	F/1O \\
ZTFJ205001.31+462426.6$^a$  & 312.5054615 &	   46.4073893 &	0.76147	&	9.26E$-$189	&	0.61109	&	1.07E$-$20	&	0.80251 &	1O/2O \\
ZTFJ205451.68+481851.5$^b$  & 313.7153451 &    48.3143193 &	1.64799	&	2.97E$-$76	&	1.18856	&	1.28E$-$45	&	0.72121 &	F/1O \\
ZTFJ210226.95+460422.8$^c$	& 315.6122978 &	   46.0730071 &	0.94812	& 3.02E$-$153	& 0.76080 & 	7.86E$-$50 &	0.80243 &	1O/2O \\
ZTFJ211100.38+480237.4$^a$  & 317.7516196 &	   48.0437494 &	1.13506	&	1.69E$-$137	&	0.91295	&	6.64E$-$33	&	0.80432 & 1O/2O \\
ZTFJ222518.95+580933.8$^a$  & 336.3289752 &	   58.1594015 &	0.99521	&	2.88E$-$160	&	0.80085	&	1.51E$-$14	&	0.80470 &	1O/2O \\
ZTFJ225650.33+622312.1$^c$  & 344.2097136 &	   62.3866983 &	1.75898	& 1.16E$-$102	& 1.27310 &	6.04E$-$23 &	0.72377 &	F/1O \\
ZTFJ230135.57+585900.6$^a$  & 345.3982421 &    58.9835105 &	1.40432	&	4.71E$-$17	&	1.01670	&	5.65E$-$14	&	0.72398 &	F/1O \\
ZTFJ230636.78+621943.6$^a$  & 346.6532848 &	   62.3288038 &	1.42788	&	4.78E$-$208	&	1.03493	&	5.69E$-$09	&	0.72480 &	F/1O \\
ZTFJ013218.14+562958.1$^b$	&  23.0756200 &	   56.4994752 &	0.87454	&	2.74E$-$124	&	0.70218	&	4.57E$-$33	&	0.80291 &	1O/2O \\
ZTFJ015700.20+573625.9$^a$	&  29.2508574 &	   57.6072142 &	1.04942	&	8.18E$-$128	&	0.84227	&	1.13E$-$66	&	0.8026 &	1O/2O \\
ZTFJ034219.03+542943.8$^a$	&  55.5793324 &	   54.4955132 &	0.89610	&	1.09E$-$108	&	0.72071	&	7.87E$-$06	&	0.80428 & 1O/2O \\
ZTFJ191145.52+120006.2$^a$	& 287.9397036 &	   12.0017394 &	1.70569	&	1.52E$-$40	&	1.36988	&	3.62E$-$07	&	0.80312 & 1O/2O \\
ZTFJ191146.64+113630.6$^c$	& 287.9443487 &	   11.6085004 &	0.41950 &	7.80E$-$154 &	0.33451 &	7.18E$-$22 &	0.79740 & 1O/2O \\
ZTFJ190319.82+152957.6$^a$	& 285.8325913 &	   15.4993410 &	0.25515	&	8.42E$-$126	&	0.20343	&	3.58E$-$17	&	0.79732 & 1O/2O \\
ZTFJ041005.43+614638.1$^a$	&  62.5226374 &	   61.7772691 &	1.11338	&	8.54E$-$76	&	0.89471	&	4.05E$-$16	&	0.80360 & 1O/2O \\
ZTFJ044523.91+425520.3$^a$	&  71.3496452 &	   42.9223088 &	0.53382	&	4.29E$-$110	&	0.42899	&	1.95E$-$39	&	0.80363 & 1O/2O \\
ZTFJ060658.08+252402.2$^b$	&  91.7420220 &	   25.4006306 &	0.61128	&	6.02E$-$84	&	0.49221	&	5.76E$-$18	&	0.80522 & 1O/2O \\
%461202300006319	& 103.5723217 &	    1.4632964 &	0.641 &	4.46E$-$107	& 0.472	& 2.36E$-$07  &	0.737 & F/1O - 2319\\
ZTFJ183335.23$-$102538.0$^b$	& 278.3968111 &	$-$10.4272427 &	6.29327	&	2.24E$-$18	&	4.38530	&	9.55E$-$20	&	0.69682 & F/1O \\
ZTFJ192801.24+195659.3$^a$	& 292.0051753 &	   19.9498102 &	4.04453	&	1.30E$-$80	&	2.80727	&	2.80E$-$75	&	0.69409 & F/1O \\
ZTFJ205127.87+461812.6$^b$	& 312.8661391 &	   46.3035242 &	3.16027	&	2.08E$-$125	&	2.23582	&	5.04E$-$56	&	0.70747 & F/1O \\
ZTFJ211839.90+504732.8$^a$	& 319.6662865 &	   50.7924570 &	2.99769	&	3.80E$-$180	&	2.11915	&	1.60E$-$24	&	0.70693 & F/1O \\
ZTFJ002537.70+641347.7$^b$	&   6.4071132 &	   64.2299167 &	3.02448	&	6.76E$-$62	&	2.15547	&	9.09E$-$65	&	0.71267 & F/1O \\
ZTFJ013859.97+645921.2$^b$	&  24.7499060 &	   64.9892500 &	2.87147	&	1.64E$-$183	&	2.03188	&	2.93E$-$14	&	0.70761 & F/1O \\
ZTFJ224743.67+573421.5$^a$	& 341.9319689 &	   57.5726424 &	2.60704	&	2.52E$-$112	&	1.85596	&	9.34E$-$217	&	0.71190 & F/1O \\
ZTFJ192549.99+194925.1$^a$  & 291.4583240 &	   19.8236574 &	3.50719 &	7.69E$-$82  & 5.03600 & 6.56E$-$89  &	0.69643 & F/1O \\
ZTFJ210811.91+460046.7$^a$  & 317.0496508 &    46.0129808 &	1.16427 &	7.82E$-$229 & 1.61687 & 3.68E$-$218 &	0.72008 & F/1O \\
ZTFJ220413.57+574316.2$^a$  & 331.0565821 &	   57.7211747 &	2.71994	& 1.38E$-$147	& 3.83044	& 1.12E$-$229 &	0.71009 & F/1O \\
ZTFJ002518.14+604553.3$^b$  &   6.3256062 &	   60.7648312 &	2.65332	& 1.43E$-$47 &	3.73469	& 1.35E$-$127 &	0.71045 & F/1O \\
ZTFJ205714.60+462338.8$^c$  & 314.3108775 &	   46.3941281 &	1.17888 &	1.31E$-$127 &	1.73681 &	1.74E$-$05 &	0.67876 & F/1O \\
ZTFJ185513.28+081813.6$^c$ & 283.8053630 & 8.3037909 & 1.86719 &  8.93E-67 &  1.34256 & 3.85E-26 & 0.71903 & F/1O \\
ZTFJ054703.02+174447.8$^a$ &86.7625850  &17.7465535  & 0.78963 & 8.54E$-$79  & 0.63560 & 1.36E$-$46 & 0.80493 & 1O/2O\\  
ZTFJ064134.57+075639.7$^b$ &100.3940561 &7.9443372   & 1.28862 & 9.20E$-$62  & 1.03145 & 8.43E$-$26 & 0.80043 & 1O/2O\\  
ZTFJ065046.49$-$085808.5$^a$ &102.6937396 &$-$8.9691016  & 2.58792 & 2.60E$-$33  & 3.60576 & 2.20E$-$47 & 0.71772 & F/1O \\  
ZTFJ071012.21$-$153204.2$^a$ &107.5509129 &$-$15.5345878 & 3.69437 & 1.13E$-$22  & 2.65381 & 1.48E$-$34 & 0.71834 & F/1O \\ 
ZTFJ072219.30$-$154455.1$^a$ &110.5805681 &$-$15.7488266 & 0.82470 & 5.22E$-$42  & 0.66402 & 1.51E$-$26 & 0.80517 & 1O/2O\\  
ZTFJ181640.89$-$105741.4$^a$ &274.1703760 &$-$10.9615102 & 8.52767 & 6.45E$-$64  & 5.91765 & 2.59E$-$58 & 0.69394 & F/1O \\  
ZTFJ183520.59$-$005344.7$^a$ &278.8358106 &$-$0.8957374  & 0.87136 & 6.40E$-$104 & 0.69966 & 1.96E$-$39 & 0.80295 & 1O/2O\\  
ZTFJ190036.66+012230.7$^a$ &285.1529969 &1.3752143   & 4.16907 & 2.20E$-$93  & 2.92370 & 3.07E$-$108& 0.70128 & F/1O \\  
\enddata
\end{deluxetable*}

\begin{deluxetable*}{ccccccccc}
\tablenum{1}
\tablecaption{(Continued.)}
\tablewidth{0pt}
\tablehead{
\colhead{ZTF ID} & \colhead{R.A. (J2000)} & \colhead{Dec. (J2000)} & \colhead{$P_1$} & \colhead{FAP1} & \colhead{$P_2$} & \colhead{FAP2} & \colhead{Period Ratio} & \colhead{Modes}\\
\colhead{} & \colhead{(deg)} & \colhead{(deg)} & \colhead{(days)} & \colhead{} & \colhead{(days)} & \colhead{} & \colhead{} & \colhead{}
}
%\decimalcolnumbers
\startdata
ZTFJ190633.59+074411.4$^c$ &286.6399870 &7.7365169   & 2.35843 & 1.32E$-$97  & 1.66720 & 3.13E$-$41 & 0.70691 & F/1O \\  
ZTFJ060249.64+184846.7$^a$ &90.7069559  &18.8128179  & 0.65314 & 3.38E$-$71  & 0.52635 & 2.47E$-$51 & 0.80588 & 1O/2O\\  
ZTFJ062805.03+142806.6$^a$ &97.0209950  &14.4684825  & 0.64072 & 1.93E$-$80  & 0.51575 & 1.41E$-$25 & 0.80495 & 1O/2O\\  
ZTFJ062855.82+110729.3$^a$ &97.2326051  &11.1248038  & 1.09668 & 2.78E$-$65  & 0.88068 & 1.03E$-$39 & 0.80304 & 1O/2O\\  
ZTFJ071659.29$-$151824.8$^b$ &109.2472005 &$-$15.3070503 & 2.31124 & 9.09E$-$26  & 1.64473 & 8.75E$-$23 & 0.71162 & F/1O \\  
ZTFJ184059.48$-$054601.4$^a$ &280.2478525 &$-$5.7670548  & 4.80809 & 2.54E$-$53  & 3.28484 & 1.24E$-$27 & 0.68319 & F/1O \\  
ZTFJ184357.04$-$024614.0$^a$ &280.9877159 &$-$2.7705098  & 1.03738 & 1.43E$-$142 & 0.82907 & 3.22E$-$56 & 0.79920 & 1O/2O\\  
ZTFJ064219.76-031854.6$^a$ &100.5824297 &$-$3.3152906  & 1.37668 & 2.02E$-$61  & 1.10499 & 4.52E$-$36 & 0.80265 & 1O/2O\\  
ZTFJ055219.11+174439.5$^a$ &88.0796577  &17.7442702  & 0.80633 & 3.60E$-$74  & 0.64934 & 5.86E$-$57 & 0.80530 & 1O/2O\\  
ZTFJ055634.35+161752.8$^a$ &89.1431964  &16.2979392  & 0.54802 & 5.47E$-$82  & 0.44165 & 7.52E$-$56 & 0.80590 & 1O/2O\\  
ZTFJ062040.28+080858.7$^a$ &95.1678464  &8.1496330   & 0.93473 & 2.04E$-$110 & 0.75066 & 1.10E$-$79 & 0.80308 & 1O/2O\\  
ZTFJ062542.06+082944.5$^a$ &96.4252827  &8.4957151   & 0.72783 & 1.61E$-$70  & 0.58714 & 7.76E$-$40 & 0.80670 & 1O/2O\\  
ZTFJ063636.38+060931.7$^a$ &99.1517198  &6.1586902   & 0.71876 & 1.02E$-$66  & 0.57720 & 9.47E$-$30 & 0.80305 & 1O/2O\\  
ZTFJ063814.13+061839.0$^a$ &99.5589704  &6.3107454   & 0.83307 & 3.72E$-$69  & 0.67093 & 4.45E$-$45 & 0.80537 & 1O/2O\\  
ZTFJ063819.23+071646.5$^a$ &99.5801659  &7.2795846   & 0.49332 & 3.46E$-$75  & 0.39761 & 3.15E$-$15 & 0.80599 & 1O/2O\\  
ZTFJ065015.78+033810.3$^a$ &102.5657596 &3.6361556   & 0.58819 & 4.67E$-$60  & 0.47334 & 1.91E$-$29 & 0.80474 & 1O/2O\\  
ZTFJ071230.19$-$165412.6$^a$ &108.1259255 &$-$16.9036312 & 0.58938 & 1.47E$-$44  & 0.47620 & 1.38E$-$23 & 0.80797 & 1O/2O\\  
ZTFJ071756.78$-$132534.9$^a$ &109.4865826 &$-$13.4264447 & 0.33334 & 1.04E$-$66  & 0.26785 & 3.69E$-$41 & 0.80353 & 1O/2O\\  
ZTFJ072206.69$-$173810.6$^a$ &110.5281129 &$-$17.6364520 & 0.67705 & 1.57E$-$38  & 0.54572 & 2.70E$-$24 & 0.80603 & 1O/2O\\  
ZTFJ185613.58$-$020059.6$^a$ &284.0566073 &$-$2.0165497  & 0.28563 & 2.12E$-$122 & 0.22656 & 2.29E$-$79 & 0.79319 & 1O/2O\\  
ZTFJ181319.70$-$163319.2$^a$ &273.3321638 &$-$16.5553899 & 7.53182 & 3.97E$-$66  & 5.16720 & 2.52E$-$40 & 0.68605 & F/1O \\  
ZTFJ073335.45$-$255036.3$^c$ &113.3979154 &$-$25.8436666 & 2.87103 & 3.79E$-$19  & 2.04826 & 8.36E$-$14 & 0.71342 & F/1O \\  
ZTFJ222304.62+574439.5$^a$ &335.7692700 &57.7443300  & 1.12015 & 6.83E$-$172 & 0.89936 & 3.43E$-$95 & 0.80290 & 1O/2O\\
\enddata
\tablecomments{ZTF ID: Source ID; R.A., Dec.: Source position (J2000); $P_1$: Dominant period obtained from the time-series data; FAP1: Corresponding FAP; $P_2$: Dominant period obtained from the light curve, pre-whitened with $P_1$; FAP2: Corresponding FAP; Period Ratio: Ratio of the shorter to the longer period; Modes: Pulsation modes of the double-mode Cepheids. $^a$ Sources detected in both the $g$ and $r$ bands and for which both $P_1$ and $P_2$ were the same in both bands; FAP values pertaining to the $g$ band are included in the table. $^b$ Sources only detected in the $g$ band; $^c$ Sources only detected in the $r$ band.}
\end{deluxetable*}

\begin{deluxetable*}{cccccccc}[h]
\tablenum{2}
\tablecaption{Double-mode Cepheids returned by our analysis and already present in the OGLE catalog\label{tab:OGLE_DMC}}
\tablewidth{0pt}
\tablehead{
\colhead{OGLE ID} & \colhead{ZTF ID} & \colhead{R.A. (J2000)} & \colhead{Dec. (J2000)} & \colhead{OGLE $P_1$} & \colhead{ZTF $P_1$} & \colhead{OGLE $P_2$} & \colhead{ZTF $P_2$} \\
\colhead{} & \colhead{} & \colhead{(deg)} & \colhead{(deg)} & \colhead{(days)} & \colhead{(days)} & \colhead{(days)} & \colhead{(days)}
}
%\decimalcolnumbers
\startdata
OGLE-GD-CEP-1275	&	ZTFJ054703.02+174447.8	&	86.762585	  &	17.7465535	&	0.78965	&	0.78963	&	0.63559	&	0.63560 \\
OGLE-GD-CEP-0057	&	ZTFJ064134.57+075639.7	&	100.3940561	&	7.9443372	  &	1.28862	&	1.28862	&	1.03152	&	1.03145 \\
OGLE-GD-CEP-1595$^a$	&	ZTFJ065046.49$-$085808.5	&	102.6937396	&	$-$8.9691016	&	3.60623	&	2.58792	&	2.58798	&	3.60576 \\
OGLE-GD-CEP-0088	&	ZTFJ071012.21$-$153204.2	&	107.5509129	&	$-$15.5345878	&	3.69488	&	3.69437	&	2.65343	&	2.65381 \\
OGLE-GD-CEP-0101	&	ZTFJ072219.30$-$154455.1	&	110.5805681	&	$-$15.7488266	&	0.82472	&	0.82470	&	0.66395	&	0.66402 \\
OGLE-BLG-CEP-152	&	ZTFJ181640.89$-$105741.4	&	274.170376	&	$-$10.9615102	&	8.52790	&	8.52767	&	5.91155	&	5.91765 \\
OGLE-GD-CEP-1376	&	ZTFJ183520.59$-$005344.7	&	278.8358106	&	$-$0.8957374	&	0.87137	&	0.87136	&	0.69970	&	0.69966 \\
OGLE-GD-CEP-1266	&	ZTFJ190036.66+012230.7	&	285.1529969	&	1.3752143	  &	4.16886	&	4.16907	&	2.92363	&	2.92370 \\
OGLE-GD-CEP-1447	&	ZTFJ190633.59+074411.4	&	286.639987	&	7.7365169	  &	2.35832	&	2.35843	&	1.66738	&	1.66720 \\
OGLE-GD-CEP-0022	&	ZTFJ060249.64+184846.7	&	90.7069559	&	18.8128179	&	0.65317	&	0.65314	&	0.52654	&	0.52635 \\
OGLE-GD-CEP-0037	&	ZTFJ062805.03+142806.6	&	97.020995	  &	14.4684825	&	0.64071	&	0.64072	&	0.51575	&	0.51575 \\
OGLE-GD-CEP-0039	&	ZTFJ062855.82+110729.3	&	97.2326051	&	11.1248038	&	1.09666	&	1.09668	&	0.88072	&	0.88068 \\
OGLE-GD-CEP-0097	&	ZTFJ071659.29$-$151824.8	&	109.2472005	&	$-$15.3070503	&	2.31109	&	2.31124	&	1.64558	&	1.64473 \\
OGLE-GD-CEP-1216	&	ZTFJ184059.48$-$054601.4	&	280.2478525	&	$-$5.7670548	&	4.80794	&	4.80809	&	3.28552	&	3.28484 \\
OGLE-GD-CEP-1226	&	ZTFJ184357.04$-$024614.0	&	280.9877159	&	$-$2.7705098	&	1.03735	&	1.03738	&	0.82909	&	0.82907 \\
OGLE-GD-CEP-1302	&	ZTFJ064219.76$-$031854.6	&	100.5824297	&	$-$3.3152906	&	1.37673	&	1.37668	&	1.10476	&	1.10499 \\
OGLE-GD-CEP-1584	&	ZTFJ055219.11+174439.5	&	88.0796577	&	17.7442702	&	0.80632	&	0.80633	&	0.64936	&	0.64934 \\ 
OGLE-GD-CEP-1585	&	ZTFJ055634.35+161752.8	&	89.1431964	&	16.2979392	&	0.54803	&	0.54802	&	0.44167	&	0.44165 \\
OGLE-GD-CEP-1587	&	ZTFJ062040.28+080858.7	&	95.1678464	&	8.149633	  &	0.93468	&	0.93473	&	0.75065	&	0.75066 \\
OGLE-GD-CEP-1588	&	ZTFJ062542.06+082944.5	&	96.4252827	&	8.4957151	  &	0.72781	&	0.72783	&	0.58714	&	0.58714 \\
OGLE-GD-CEP-1591	&	ZTFJ063636.38+060931.7	&	99.1517198	&	6.1586902	  &	0.71880	&	0.71876	&	0.57709	&	0.57720 \\
OGLE-GD-CEP-1592	&	ZTFJ063814.13+061839.0	&	99.5589704	&	6.3107454	  &	0.83306	&	0.83307	&	0.67106	&	0.67093 \\
OGLE-GD-CEP-1593	&	ZTFJ063819.23+071646.5	&	99.5801659	&	7.2795846	  &	0.49332	&	0.49332	&	0.39760	&	0.39761 \\
OGLE-GD-CEP-1594	&	ZTFJ065015.78+033810.3	&	102.5657596	&	3.6361556	  &	0.58812	&	0.58819	&	0.47309	&	0.47334 \\
OGLE-GD-CEP-1598	&	ZTFJ071230.19$-$165412.6	&	108.1259255	&	$-$16.9036312	&	0.58939	&	0.58938	&	0.47619	&	0.47620 \\
OGLE-GD-CEP-1599	&	ZTFJ071756.78$-$132534.9	&	109.4865826	&	$-$13.4264447	&	0.33334	&	0.33334	&	0.26788	&	0.26785 \\
OGLE-GD-CEP-1602	&	ZTFJ072206.69$-$173810.6	&	110.5281129	&	$-$17.636452	&	0.67703	&	0.67705	&	0.54572	&	0.54572 \\
OGLE-GD-CEP-1803	&	ZTFJ185613.58$-$020059.6	&	284.0566073	&	$-$2.0165497	&	0.28563	&	0.28563	&	0.22656	&	0.22656 \\
OGLE-BLG-CEP-147  & ZTFJ181319.70$-$163319.2  & 273.3321638 & $-$16.5553899 & 7.53251 & 7.53182 & 5.16492 & 5.16720 \\
OGLE-GD-CEP-0117  & ZTFJ073335.45$-$255036.3  & 113.3979154 & $-$25.8436666 & 2.87088 & 2.87103 & 2.04830 & 2.04826 \\
\enddata
\tablecomments{OGLE, ZTF ID: IDs in the OGLE and ZTF catalogs, respectively; R.A., Dec.: Source position (J2000); OGLE $P_1$, ZTF $P_1$: Dominant periods listed in the OGLE catalog and returned by our analysis of ZTF DR5 data, respectively; OGLE $P_2$, ZTF $P_2$: Secondary periods listed in the OGLE catalog and returned by our analysis of ZTF DR5 data, respectively.\\
$^a$ Note that the ZTF $P_1$ and $P_2$ values determined for this source are equivalent to, respectively, the OGLE $P_2$ and $P_1$ values.}
\end{deluxetable*}

\section{Discussion} \label{sec:disc}

\begin{figure}
	    \centering
	    \includegraphics[scale=0.8]{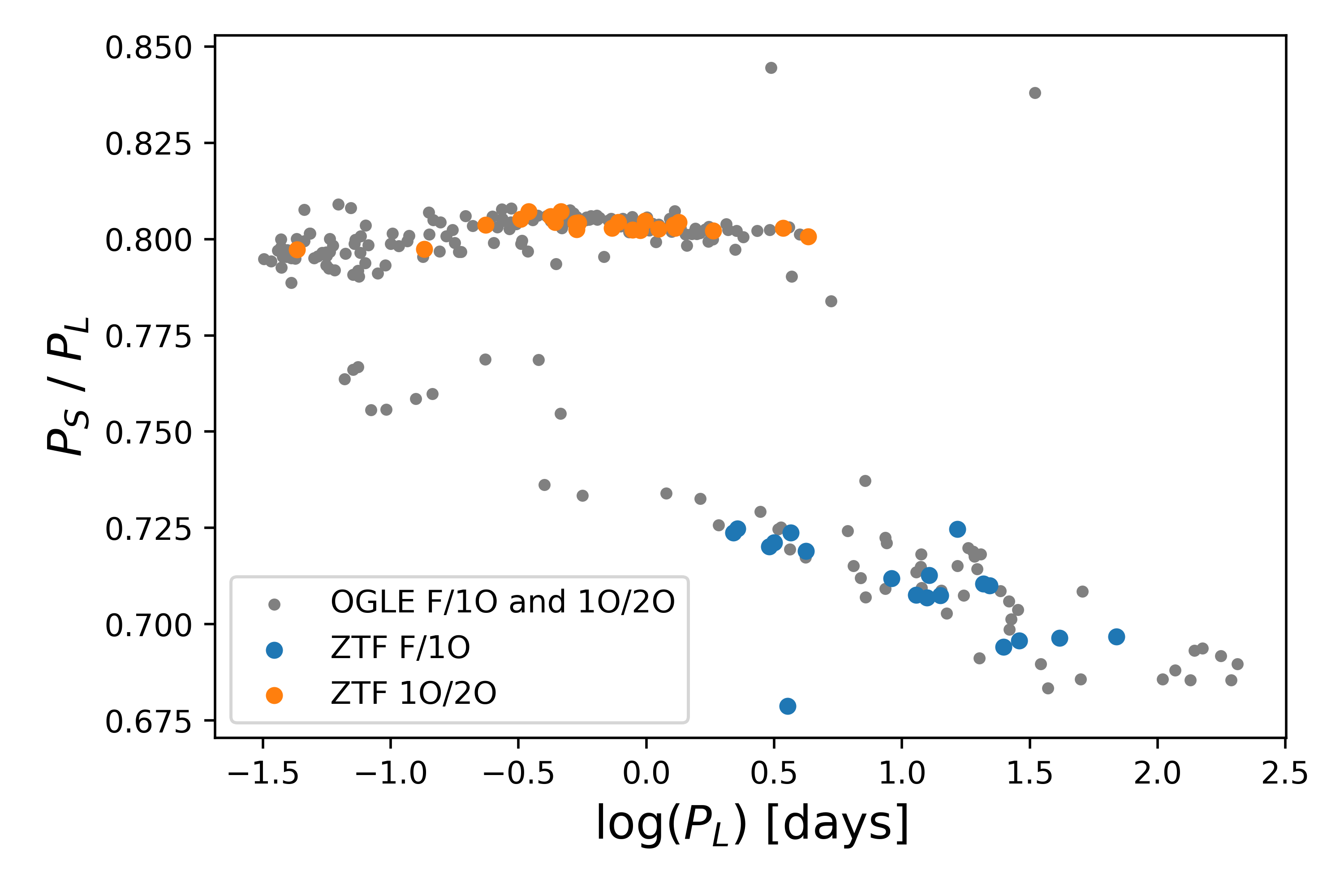}
	    \caption{Petersen diagram representing the distribution of OGLE and ZTF F/1O and 1O/2O Cepheids. $P_{\rm S}$ and $P_{\rm L}$ represent the shorter and longer periods, respectively. The F/1O and 1O/2O Cepheids not included in the OGLE catalog are shown as blue and orange dots, respectively. The OGLE F/1O and 1O/2O Cepheids are indicated by gray dots.}\label{fig:3}
\end{figure}

Figure \ref{fig:3} shows our new double-mode Cepheids (those not included in the OGLE database) in the Petersen diagram, along with the Galactic double-mode Cepheids from OGLE. $P_{\rm S}/P_{\rm L}$ represents the ratio of the shorter and longer periods, whereas $\log( P_{\rm L})$ is the logarithm of the longer periods. The loci of our ZTF Cepheids are in excellent agreement with the trend followed by the OGLE Cepheids. One of our double-mode Cepheid candidates, with $P_{\rm S}/P_{\rm L} = 0.67876, P_1 = 1.17888$ days, and $P_2 = 1.73681$ days is located outside the expected F/1O locus in the Petersen diagram. This object might be a 1O Cepheid exhibiting additional low-amplitude periodicity \citep[e.g.,][]{2020past.conf...75Z}. The diagram shows that all of our double-mode Cepheids are found toward the higher end of the $P_{\rm L}$ range. This is likely driven by the selection criteria adopted by \citet{paper1}, which particularly affects the reliability of our selection of short-period Cepheids ($P_1 < 1$ day). In the absence of luminosity information, these candidate Cepheids may be contaminated by other types of short-period variables given the similarities in their light curves at these short periods and in view of the prevailing parallax uncertainties.

Among the double-mode Cepheids listed in Table \ref{tab:new_DMC}, six have $P_1 < P_2$, including ZTFJ192549.99+194925.1, ZTFJ210811.91+460046.7, ZTFJ220413.57+574316.2, ZTFJ002518.14+604553.3, ZTFJ065046.49$-$085808.5, and ZTFJ205714.60+462338.8. For these Cepheids, the first-overtone frequencies are associated with larger amplitudes, whereas their fundamental frequencies are suppressed.

Our initial sample included 42 of OGLE's Galactic F/1O and 1O/2O Cepheids. Of these, we identified 30 as possible double-mode Cepheids, thus suggesting that our method is 71\% efficient at finding double-mode Cepheids. We analyzed the remaining 12 sources individually to determine why they were not identified by our analysis. Table \ref{tab:NOT_IDEN} lists the OGLE periods of these sources, as well as the ZTF periods and FAPs returned by our analysis, and the periods and FAPs returned by running our code on the OGLE data of these sources. As is evident from the table, our analysis did not identify these candidates because the FAP values based on the ZTF data are $>$ 0.001. Based on the OGLE data, our code can identify these double-mode Cepheids, but their FAPs are mostly greater than $10^{-10}$. This indicates that the secondary periods of these double-mode Cepheids are not significant, and the number of ZTF photometric data points is not sufficient to detect these secondary periods. 

\movetabledown=1.5cm
\begin{longrotatetable}
\begin{deluxetable*}{cccccccccccc}
\tablenum{3}
\tablecaption{OGLE double-mode Cepheids in our sample of classical Cepheids but not identified by our code\label{tab:NOT_IDEN}}
%\tablewidth{500pt}
\tablehead{
\colhead{OGLE ID} & \colhead{ZTF ID} & \colhead{ZTF $P_1$} & \colhead{ZTF} & \colhead{OGLE $P_1$} & \colhead{OGLE} & \colhead{OGLE $P_1$} &  \colhead{ZTF $P_2$} & \colhead{ZTF} & \colhead{OGLE $P_2$} & \colhead{OGLE} & \colhead{OGLE $P_2$} \\
\colhead{} & \colhead{} & \colhead{} & \colhead{FAP1} & \colhead{} & \colhead{FAP1} & \colhead{(catalog)} &  \colhead{} & \colhead{FAP2} & \colhead{} & \colhead{FAP2} & \colhead{(catalog)} \\
\colhead{} & \colhead{} & \colhead{(days)} & \colhead{} & \colhead{(days)} & \colhead{} & \colhead{(days)} & \colhead{(days)} & \colhead{} & \colhead{(days)} & \colhead{} & \colhead{(days)}
}
%\decimalcolnumbers
\startdata
OGLE-GD-CEP-0064 & ZTFJ064545.66$-$035146.5 & 1.83988 & 4.19E$-$92 & 1.83988 & 2.28E$-$107 & 2.54647 & 1.40638 & 9.09E$-$01 & 1.26430 & 9.83E$-$01	&	1.83987 \\
OGLE-GD-CEP-0106 & ZTFJ072859.96$-$231134.9 & 0.28976 & 1.02E$-$37 & 0.28976 & 3.57E$-$106 & 0.28976 & 0.19960 & 2.44E$-$02 & 0.23184 & 3.34E$-$06	&	0.23184 \\
OGLE-GD-CEP-0139 & ZTFJ075308.92$-$260300.5 & 0.58759 & 2.49E$-$17 & 0.58758 & 1.32E$-$111 & 0.58758 & 0.41502 & 1.56E$-$01 & 0.47246 & 1.42E$-$17	&	0.47247 \\
OGLE-GD-CEP-1604 & ZTFJ072512.92$-$210444.2 & 3.89821 & 7.07E$-$38 & 3.89919 & 1.84E$-$95  & 5.50298 & 2.95551 & 2.68E$-$01 & 2.97555 & 9.68E$-$01	&	3.89932 \\
OGLE-GD-CEP-1612 & ZTFJ074757.30$-$274927.5 & 0.23966 & 2.08E$-$32 & 0.23967 & 2.13E$-$95  & 0.23966 & 0.19895 & 3.91E$-$02 & 0.19171 & 6.47E$-$22	&	0.19171 \\
OGLE-GD-CEP-1788 & ZTFJ182833.76$-$102742.1 & 0.22392 & 2.83E$-$68 & 0.22392 & 6.52E$-$36  & 0.22392 & 0.16534 & 2.29E$-$02 & 0.17806 & 1.15E$-$05	&	0.17799 \\
OGLE-GD-CEP-1790 & ZTFJ183258.00$-$104302.8 & 0.24925 & 3.69E$-$19 & 0.24925 & 5.34E$-$23  & 0.24925 & 0.16409 & 3.75E$-$02 & 0.19817 & 1.69E$-$03	&	0.19817 \\
OGLE-GD-CEP-1797 & ZTFJ184711.94$-$090312.4 & 4.08245 & 4.07E$-$02 & 0.29994 & 2.91E$-$17  & 0.29994 & 3.07714 & 1.00E+00 & 0.24266 & 1.45E$-$03	&	0.24266 \\
OGLE-GD-CEP-1798 & ZTFJ184735.17$-$013535.0 & 0.61226 & 5.28E$-$54 & 0.61228 & 7.45E$-$49  & 0.61229 & 0.48913 & 1.47E$-$01 & 0.48912 & 5.22E$-$10	&	0.48912 \\
OGLE-GD-CEP-1804 & ZTFJ185958.20$-$014220.9 & 0.24109 & 2.48E$-$22 & 0.24109 & 2.05E$-$48  & 0.24109 & 0.15963 & 3.48E$-$01 & 0.19178 & 3.31E$-$08	&	0.19178 \\
OGLE-BLG-CEP-077 & ZTFJ175045.97$-$225949.0 & 0.23420 & 3.09E$-$03 & 3.51965 & 3.03E$-$87  & 3.51969 & 0.17751 & 1.00E+00 & 2.53351 & 3.20E$-$215 &	2.5330  \\
OGLE-GD-CEP-1799 & ZTFJ185322.20$-$085832.8 & 0.45989 & 6.81E$-$20 & 0.45987 & 5.53E$-$21  & 0.31474 & 0.34139 & 4.64E$-$31 & 0.34139 & 1.36E$-$29  & 0.25433 \\
\enddata
\tablecomments{OGLE, ZTF ID: IDs in the OGLE and ZTF catalogs, respectively; ZTF $P_1$, $P_2$: Period for the ZTF data; ZTF FAP1, FAP2: Corresponding FAP; OGLE $P_1$, $P_2$: Period obtained by running our code on the OGLE data; OGLE FAP1, FAP2: Corresponding FAP; OGLE $P_1$, $P_2$ (catalog): Period listed in the OGLE catalog.}
\end{deluxetable*}
\end{longrotatetable}

\section{Conclusion}
\label{sec:concl}
Our analysis of the ZTF DR5 data has identified 72 Galactic double-mode Cepheids (29 F/1O and 43 1O/2O). The loci of these Cepheids in the Petersen diagram agree with the general trend followed by double-mode Cepheids, further confirming their correct classification. Thirty double-mode Cepheids were already listed in the OGLE catalog. Our periods are in excellent agreement with the periods reported by OGLE. Twelve additional OGLE double-mode Cepheids were present in our sample but not identified by our analysis. The main reason for this is that the secondary periods of these double-mode Cepheids are of low significance and the number of ZTF photometric data points is not sufficient to detect them. Compared with OGLE, our completeness is around 71\%. With better sampling of light curves in future ZTF DRs, we expect to identify an even larger number of new Galactic double-mode Cepheids, which will be complementary to the OGLE Galactic double-mode Cepheids in the northern sky.

\section*{Acknowledgements}
X.C. acknowledges funding support from the National Natural Science Foundation of China (NSFC) through grants 12173047 and 11903045. This research was supported in part by the Australian Research Council Centre of Excellence for All Sky Astrophysics in 3 Dimensions (ASTRO 3D), through project number CE170100013. This publication is based on observations obtained with the Samuel Oschin 48-inch Telescope at Palomar Observatory as part of the Zwicky Transient Facility project. ZTF is supported by the U.S. National Science Foundation through grant AST-1440341 and a collaboration including Caltech, IPAC, the Weizmann Institute for Science, the Oskar Klein Center at Stockholm University, the University of Maryland, the University of Washington, Deutsches Elektronen-Synchrotron and Humboldt University, Los Alamos National Laboratories, the TANGO Consortium of Taiwan, the University of Wisconsin at Milwaukee and Lawrence Berkeley National Laboratories. Operations are conducted by Caltech Optical Observatories, IPAC and the University of Washington.

%\renewcommand{\bibname}{References}
%\bibliographystyle{unsrt}
%\bibliographystyle{aasjournal}
%\bibliography{DMCs}

\end{document}